\documentstyle[12pt]{article}
\begin{document}

\begin{flushright}
preprint UT-674\\
April, 1994 \\
hep-th/9404056
\end{flushright}

\bigskip

\begin{center}
Toda Lattice Hierarchy and the Topological
\\
Description of the $c=1$ String Theory

          {~\\~\\~\\
          {\it Tohru Eguchi} \\
          {\it Department of Physics, Faculty of Science}\\
          {\it University of Tokyo} \\
          {\it Tokyo 113, Japan} \\
          ~\\
           {\it and} \\
          ~\\
          {\it Hiroaki Kanno} \\
          {\it Department of Mathematics} \\
          {\it Hiroshima University}\\
          {\it Higashi-Hiroshima 724, Japan}}

\end{center}

\newcommand {\beq}{\begin{equation}}
\newcommand {\eeq}{\end{equation}}
\newcommand {\bea}{\begin{eqnarray}}
\newcommand {\eea}{\end{eqnarray}}

\vspace{10mm}

\begin{abstract}
The Toda lattice hierarchy is discussed in connection with the topological
description of the $c=1$ string theory compactified at the self-dual radius.
It is shown that when special constraints are imposed on the Toda hierarchy,
it reproduces known results of the  $c=1$ string theory, in particular the
$W_{1+\infty}$ relations among tachyon correlation functions. These constraints
are the analogues of string equations in the topological minimal theories.
We also point out that at $c=1$ the Landau-Ginzburg superpotential
becomes simply a $U(1)$ current operator.
\end{abstract}
\newpage

Recently a topological description of the $c=1$ string theory has been
proposed in \cite{HOP,GM} using the Landau-Ginzburg formulation. In these
articles tachyon
correlation functions of the $c=1$ theory compactified at the self-dual radius
are shown to be reproduced by the topological Landau-Ginzburg theory with a
superpotential
of the form $W=1/x+ \cdots$. This suggests that the $c=1$ theory may be
identified as the
$k=-3$ extension of the topological minimal theories with the superpotential
$W=x^{k+2}+\cdots$ as
discussed in \cite{W,MV}. In refs.
\cite{HOP,GM} positive momentum tachyons $T_n  (n=1,2,\cdots)$ with
momenta $n$ (in the unit of $1/\sqrt{2}$) are regarded as primary fields and
$T_1$
is identified as the puncture operator. On the other hand
negative momentum tachyons $\bar{T}_n  (n=1,2,\cdots)$ with momenta
$-n$ are regarded as the descendants of the cosmological constant operator
$T_0$.
Some irregular behaviors of the $N$-point tachyon correlation functions \cite
{DK} are
interpreted as being due to the existence of contact terms among
the gravitational descendants.

 On the other hand, using the results of the S-matrix calculation of the $c=1$
matrix model \cite{MPR,P,MP} it was shown \cite{DMP} that the free energy of
the
$c=1$ theory at the self-dual radius is a tau function of the Toda lattice
hierarchy \cite{UT}. Thus the Toda lattice hierarchy may be considered as
the underlying integrable structure in the $c=1$ theory.
In this article we would like to discuss in some detail
how the Toda lattice hierarchy fits into the Landau-Ginzburg description of
the $c=1$ model.

We first point out that the integrable structure of the $c=1$ string
theory is described by the Toda lattice hierarchy when special
constraints are imposed on the Toda system (see eqs.(24),(25) below).
These constraints are the
analogues of the string equations in the minimal theories which select
special solutions of the KP hierarchy. Our conditions reproduce the
$W_{1+\infty}$
constraints discovered in \cite{MP,DMP}. It turns out that the constrained
Toda hierarchies are described
in terms of a pair of Lax-like operators $L,W$ which are the analogues of the
$P,Q$ operators \cite{D} of the $c<1$ minimal theories.
In the Landau-Ginzburg description one may regard $W(L)$ as the
superpotential and $L(W)$ as the Landau-Ginzburg field. $W$ has an expansion
in terms of $L$ and has the form
of a $U(1)$ current $W=\partial \phi/\partial L$ where $\phi(L)$
is a free scalar field $\phi=\mu\log L + \sum t_n L^n -\sum
\partial/n\partial t_n L^{-n}$.
Here $t_n$ is the coupling constant of the positive
tachyon field $T_n$ and $\mu$ is the cosmological constant
($\partial/\partial t_n$ acts on the partition function and is replaced by
the 1-point function $<T_n>$ at genus=0).
Similarly the operator $L$ is
expanded in terms of $W$ and has the form of a $U(1)$ current. From these
expressions $W_{1+\infty}$ symmetry follows immediately. We note that in the
$c=1$ theory
$L$ and $W$ plays a completely symmetric role unlike the $P$ and $Q$
operators
in the minimal theories. We also note that the free boson $\phi$ has a zero
mode
whose eigenvalue is the cosmological constant $\mu$. In the case of minimal
theories the free boson was twisted and devoid of
the zero mode \cite{DVV1,FKN}.

Let us first recall the structure of the Toda lattice hierarchy \cite{UT}.
In this paper we essentially restrict ourselves to the
genus-zero case and consider the dispersionless limit of the Toda hierarchy
\cite{TT}. One introduces two pairs of Lax-like operators, $L,M$ and $\bar{L},
\bar{M}$,
\bea
&&L=x+\sum_{i=0}^{\infty}u_ix^{-i}~,
\\
&&M=\mu+\sum_{n=1}nt_n L^n+\sum_{n=1}v_n L^{-n}~,
\\
&&\bar{L}=\sum_{i=1}^{\infty}\bar{u}_ix^i~,
\\
&&\bar{M}=\mu+\sum_{n=1}n\bar{t}_n\bar{L}^{-n}
+\sum_{n=1}\bar{v}_n\bar{L}^n~.
\eea
They obey the Poisson bracket relations
\bea
&&\{L,M\}=L~,
\\
&&\{\bar{L},\bar{M}\}=\bar{L}~,
\eea
where the Poisson bracket is defined by
\begin{equation}
\{A(x,\mu),B(x,\mu)\}=x\Big(\frac{\partial A}{\partial x}\frac{\partial
B}{\partial \mu}-
\frac{\partial A}{\partial \mu}\frac{\partial B}{\partial x}\Big)~.
\end{equation}
Time evolutions of the operators are given by
\bea
&&\frac{\partial}{\partial t_n}*\!*=\{H_n,**\}~,~~~~~~~~~~~H_n=(L^n)_+~,
{}~n=1,2,\cdots,
\\
&&\frac{\partial}{\partial \bar{t}_n}*\!*=-\{\bar{H}_n,**\}~,~~~~~~~~\bar{H}_n=
(\bar{L}^{-n})_-~,~~n=1,2,\cdots,
\eea
where $**$ stands for $L,M,\bar{L}$ and $\bar{M}$.
$(**)_+ \big((**)_-\big)$ means to take terms with non-negative (negative)
powers of $x$ in the series $(**)$. It is easy to check that the
flows (8),(9) commute with each other.

The operator $M$ is reexpressed as
\beq
M=K\big(\mu + \sum_{n=1}n t_n x^n \big)K^{-1}~,
\eeq
if one introduces the ``dressing operator'' $K$ defined by
\beq
L=KxK^{-1}~.
\eeq
In fact one can check
\beq
\{L,M\}=K\{x,\mu+\sum n t_n x^n \}K^{-1}=KxK^{-1}=L~.
\eeq
Furthermore, using $\partial K/\partial t_m =H_mK-Kx^m,\partial K/
\partial\bar{t}_m =-\bar{H}_mK$, one finds
\bea
&& \frac{\partial M}{\partial t_m}=\{H_m,M\}-K\{x^m,\mu+\sum nt_nx^n\}K^{-1}
+Kmx^mK^{-1}
\\
&&~~~~~~~=\{H_m,M\},~~~~~~~~~~~~~~~~m=1,2,\cdots
\\
&&\frac{\partial M}{\partial \bar{t}_m}=-\{\bar{H}_m,M\}~,
{}~~~~~~~~~~~~~~m=1,2,\cdots~.
\eea
Defining the coefficient functions $\{v_n\}$ by the expansion
\beq
K\mu K^{-1}=\mu+\sum_{n=1}v_nL^{-n}
\eeq
we recover the formula eq.(2). Eq.(4) may be analyzed in a similar manner.
Eqs.(1)-(4) appear to be a pair of KP systems,
however, the presence of the zero-mode $\mu$ in $M,\bar{M}$ and the extra
factor of $x$ in the RHS of eq.(7) make an important distinction between Toda
and KP systems.

It is known that the coefficient functions $\{v_n\}(\{\bar{v}_n\})$ are
$\partial/\partial t_n (\partial/\partial \bar{t}_n)$ derivatives of the
(logarithm of) tau function and hence are the positive (negative)
tachyon one-point functions
\cite{TT}
\beq
v_n=\frac{1}{Z}\frac{\partial Z}{\partial t_n}=<T_n>~,
{}~~~\bar{v}_n=\frac{1}{Z}\frac{\partial Z}{\partial \bar{t}_n}=<\bar{T}_n>.
\eeq
On the other hand, from the Poisson bracket relations (5)(6) the coefficient
functions $\{u_n\},\{\bar{u}_n\}$ are expressed in terms of
$\{v_n\},\{\bar{v}_n\}$ and their $\mu$-derivatives. Then the flow equations
(8)(9) lead to
the following relations \cite{HOP}
\bea
&&\frac{\partial v_n}{\partial t_1}=<T_1T_n>=\oint L^n dx~,
\\
&&\frac{\partial \bar{v}_n}{\partial t_1}=<T_1\bar{T}_n>=\oint \bar{L}^{-n}
dx~.
\eea
Also
\bea
&&\frac{\partial v_n}{\partial \mu}=<T_0T_n>=\oint x^{-1}L^n dx~,
\\
&&\frac{\partial \bar{v}_n}{\partial \mu}=<T_0\bar{T}_n>=\oint x^{-1}
\bar{L}^{-n} dx~
\eea
hold. We have obtained formulas for the $\bar{t}_1$-derivatives
\bea
&&\frac{\partial v_n}{\partial \bar{t}_1}=<\bar{T}_1T_n>=\frac{1}
{\bar{u}_1}\oint x^{-2}L^n dx~,
\\
&&\frac{\partial \bar{v}_n}{\partial \bar{t}_1}=<\bar{T}_1\bar{T}_n>=\frac{1}
{\bar{u}_1}\oint x^{-2}\bar{L}^{-n} dx~.
\eea
When we switch off the couplings $\{\bar{t}_n\}(\{t_n\})$, functions $\{u_n\},
\{v_n\}$ ($\{\bar{u}_n\},\{\bar{v}_n\}$) vanish. Thus we may regard $L$
as being the deformation of the variable $x$ in the presence of nonzero
descendant couplings. We identify $L$ as the Landau-Ginzburg field in the
large phase space. This is reminiscent of the work \cite{LP} where the operator
$Q$ is regarded as the Landau-Ginzburg field in the large phase space of
minimal theories.

Let us next introduce the following constraints on the Lax-like operators
\bea
&&ML^{-1}=\bar{L}^{-1}~,
\\
&&\bar{M}\bar{L}=L~.
\eea
It turns out that these conditions are compatible with the Toda system (1)-(9)
and reproduce known results of the $c=1$ string theory compactified at the
$SU(2)$ radius. We note that (24),(25)
eliminate two out of four operators $L,M,\bar{L},\bar{M}$. Let us define
\beq
W\equiv ML^{-1}=\mu L^{-1}+\sum_{n=1}nt_nL^{n-1}+\sum_{n=1}v_nL^{-n-1}
\eeq
which obeys
\beq
\{L,W\}=1~.
\eeq
We now describe the Toda hierarchy under the constraints (24),(25) by a
pair of operators $L$ and $W$. Let us first note that since $\bar{L}$ has a
Taylor expansion in $x$ starting from $x^1$ (eq.(3)), eq.(24)
implies the following expansion of W
\beq
W=a_{-1}x^{-1}+a_0+\sum_{n=1}a_n x^n~.
\eeq
Thus the operator $W$ does not contain terms $x^m$ with powers $m\leq -2$.
On the other
hand, $L$ has the expansion eq.(1) and hence $L^{-n-1} (n=1,2\cdots)$
contain only terms $x^m$ with $m\leq -2$. Therefore the last sum in
eq.(26) actually does not contribute and (26) may be
reexpressed as
\bea
&&W=a_{-1}x^{-1}+\sum nt_n(L^{n-1})_+~,
\\
&&a_{-1}=\mu+\sum nt_n(L^{n-1})_{-1}~.
\eea
Here $(**)_n$ denotes the coefficient of the $x^n$ term in $**$.

On the other hand, the constraint (25) leads to an expansion of $L$ in terms
of
$W$
\beq
L=\mu W^{-1}+\sum n\bar{t}_n W^{n-1}+\sum \bar{v}_n W^{-n-1}~.
\eeq
Again by comparing the $x$-expansion of both sides of eq.(31) it may be
reexpressed as
\bea
&&L=u_{-1} x+u_0+\sum n\bar{t}_n(W^{n-1})_-~,
\\
&&u_{-1}=\mu(a_{-1})^{-1}+\sum n\bar{t}_n(W^{n-1})_{+1}~,
\\
&&u_0=\sum n\bar{t}_n(W^{n-1})_0~.
\eea
We shall show below (see eq.(41)) that $u_{-1}=1$
and hence (32) is in fact in accord with the expansion (1). One may also
prove that (34) holds.

Now let us come to the discussions of the $W_{1+\infty}$ constraints.
Using (31) we can express the coefficient function $\bar{v}_n$ by a residue
integral
\beq
\bar{v}_n=\oint L W^n dW=\frac{1}{n+1}\oint W^{n+1} dL~,~~~n=1,2,\cdots~.
\eeq
(The direction of the integration contour flips under the change of variable
and this cancels the $-$ sign coming from the partial integration).
Similarly from (26) we have
\beq
v_n=\frac{1}{n+1}\oint L^{n+1} dW~,~~~n=1,2,\cdots~.
\eeq
If we substitute (26) into the RHS of (35), we obtain a relation representing
$\bar{v}_n$ in terms of $\{v_m\},\{t_m\}$. These are the (zero-genus version
of)
$W_{1+\infty}$ constraints \cite{MP}. Eq.(35) was postulated in \cite{HOP}
based on an analogy with the representation of 1-point functions in the minimal
theories by means of period integrals \cite{EYY}.
Eq.(36) gives the $\bar{W}_{1+\infty}$ constraints.

Now we prove $u_{-1}=1$. First by putting $n=1$ in (35),(36) we have
\bea
&&\bar{v}_1=\mu t_1+\sum_{n=2} nt_nv_{n-1}~,
\\
&&v_1=\mu\bar{t}_1+\sum_{n=2} n\bar{t}_n\bar{v}_{n-1}~.
\eea
It turns out that these formulas correspond to the puncture equation in the
case of minimal models.
Taking the $t_1$-derivative of (37) and using (18), one finds
\bea
&&\frac{\partial \bar{v}_1}{\partial t_1}=\mu+\sum_{n=2} n t_n
\frac{\partial v_{n-1}}{\partial t_1}=\mu+\sum_{n=2} n t_n(L^{n-1})_{-1}
\nonumber \\
&&~~~~~~=a_{-1}~.
\eea
On the other hand eq.(23) reads (note $a_{-1}=\bar{u}_1^{-1}$)
\beq
(W^{n-1})_{+1}=\frac{1}{a_{-1}}\frac{\partial \bar{v}_{n-1}}
{\partial \bar{t}_1}.
\eeq
Therefore
\bea
&&u_{-1}=\mu(a_{-1})^{-1}+\sum n \bar{t}_n(W^{n-1})_{+1}
\nonumber \\
&&=\mu(a_{-1})^{-1}+(a_{-1})^{-1}\frac{\partial}{\partial \bar{t}_1}
\sum n\bar{t}_n\bar{v}_{n-1}
\nonumber \\
&&=(a_{-1})^{-1}(\mu+\frac{\partial}{\partial \bar{t}_1}(v_1-\mu\bar{t}_1))=
(\frac{\partial \bar{v}_1}{\partial t_1})^{-1}\frac{\partial v_1}
{\partial \bar{t}_1}=1
\eea
where we used (38) in going from the 2nd to the 3rd line.
Thus in fact $u_{-1}=1$ and the constraints (24)(25) are fully consistent with
the Toda lattice hierarchy.

Let us now derive the analogue of the puncture equation (genus-zero version of
the string equation).
We first note that formulas (29)(32) may be expressed in an appealing manner
as a sum of the Hamiltonians (8)(9)
\bea
&&W=\bar{H}_1+\sum nt_nH_{n-1}~,
\\
&&L=H_1+\sum n\bar{t}_n\bar{H}_{n-1}~.
\eea
If we evaluate $\{L,W\}=1$ using these formulas, we obtain puncture equations
for our system
\bea
&&\frac{\partial W}{\partial t_1}-\sum n \bar{t}_n \frac{\partial W}
{\partial \bar{t}_{n-1}}=1~,
\\
&&
\frac{\partial L}{\partial \bar{t}_1}-\sum n t_n \frac{\partial L}
{\partial t_{n-1}}=1~.
\eea
It is possible to show that these formulas are equivalent to the $W^{(2)}$-
constraints eqs.(37)(38). We may, for instance, multiply (45) by
$mL^{m-1}x^{-1}$
and take the residue in $x$ on both sides of the equation.
Then by using the relation (20) we obtain
\beq
\frac{\partial}{\partial \mu}\frac{\partial v_m}{\partial \bar{t}_1}=
\frac{\partial}{\partial \mu}(\sum_{n=1}nt_n\frac{\partial v_m}
{\partial t_{n-1}}+mv_{m-1})~
\eeq
(we formally set $v_0=\mu$).
Integrating over $\mu$ (we assume the absence of the integration constant)
we obtain
\beq
\frac{\partial v_m}{\partial \bar{t}_1}=\sum_{n=1}nt_n\frac{\partial v_m}
{\partial t_{n-1}}+mv_{m-1}=\frac{\partial}{\partial t_m}
(\sum_{n=1}nt_nv_{n-1})~.
\eeq
It is easy to see that this equation agrees with the
$\partial/\partial t_m$-derivative of (37).
Hence $W^{(2)}$-constraint corresponds to the punture ($L_{-1}$) equation in
the minimal models. In fact the identification $T_1$=puncture operator
was made previously in \cite{K} via the analysis of this constraint.

Let us next discuss some formal aspects of the $W_{1+\infty}$ symmetry.
We first note that since the coefficient functions $\{v_n\},\{\bar{v}_n\}$
are 1-point functions, we may replace them by the drivatives
$\{\partial/\partial t_n\},\{\partial /\partial \bar{t}_n\}$ acting on the
partition function. Thus we can rewrite the superpotential as the $U(1)$
current
\beq
W=\partial \phi~,~~~~~~\phi=\mu \log L+\sum t_n L^n-\sum\frac{1}{n}\frac
{\partial}{\partial t_n}L^{-n}~.
\eeq
The $W_{1+\infty}$ relation (35) may then be rewritten as
\beq
\frac{1}{Z}\frac{\partial Z}{\partial \bar{t}_n}=\frac{1}{n+1}
\frac{1}{Z}\oint dL (\partial \phi)^{n+1} Z~.
\eeq
(49) reduces to (35) at genus $g=0$ since genus-zero contibutions come only
when the derivative $\{\partial/\partial t_n\}$ hits the partition function
(the genus-dependence of our system
will be recovered if we rescale the derivative term in (48) as $-\lambda^2
\sum 1/n\partial/\partial t_nL^{-n}$ and expand the free energy
as $\log Z=\sum_g \lambda^{2g-2}F_g$ where $\lambda$ is the genus-expansion
parameter
and $F_g$ is the genus-$g$ free-energy).
The $n$-th power of the $U(1)$ current $(\partial\phi)^n$
describes the spin-$n$ field $W^{(n)}$
of the $W_{1+\infty}$ algebra.
The contour integral $\oint dL (\partial \phi)^{n+1}$ extracts the mode
$W^{(n+1)}_{-n}$. In Eq.(49) the $W_{1+\infty}$ symmetry of the theory is
seen explicitly.
Thus it seems natural to regard the superpotential $W$ as the $U(1)$ current
operator in the $c=1$ theory. We clearly see how the Landau-Ginzburg
description emerges at the $g=0$
limit where the $U(1)$ operator is replaced by its expectation value.

One of the basic differences between the
Toda and KP systems is in the symmetry between their Lax-like operators.
In the case of the KP hierarchy the
$Q$ operator can be expanded in terms of $P$ \cite{D}, however,
there does not seem to be an inverse expansion of $P$ in terms of $Q$.
On the other hand, in the Toda theory there exists a complete symmetry between
$L$ and $W$ operators and they are mutaually expanded into each other.
This symmetry controls the Toda system in an efficient manner and reduces
it almost to that of free fields in the target space. In this
respect, despite the
presence of an infinity of primary fields, the $c=1$ theory is somewhat
simpler than the minimal theories.

In this paper we have restricted ourselves to the $c=1$ string theory at the
self-dual radius. It will be extremely interesting to study the theory
in different backgrounds, for instance, the 2-dimensional black hole
(for a recent attempt, see \cite{Da}). It is a challenging problem to see
if other reductions of the Toda theory are possible than the case treated in
this paper.

\bigskip

During the completion of this manuscript a new preprint by K.Takasaki \cite{T}
has appeared. This paper discusses subjects closely related to ours.

\bigskip

T.E. would like to thank A.Hanany and Y.Oz for discussions during his stay at
Tel-Aviv University in December 1993. H.K. thanks K.Takasaki for discussions.
Research of T.E. is partly supported by
the grant-in-aid for scientific research on priority area
``Infinite Analysis'', Japan Ministry of Education.

\end{document}